\pdfoutput=1
\documentclass[article,preprint,superscriptaddress]{revtex4-1}
\usepackage{graphicx}
\usepackage{dcolumn}
\usepackage{amssymb}
\usepackage{amsmath}
\usepackage{bm}
\usepackage{pifont}
\usepackage{epsfig}
\usepackage{psfrag}
\usepackage[usenames]{xcolor}

\begin{document}

\title[Influence of an amorphous wall on localized excitations]{Influence of an amorphous wall on the distribution of localized excitations in a colloidal glass-forming liquid}
\author{Shreyas Gokhale$^*,$\footnote{Present address: Physics of Living Systems group, MIT, 400 Technology Square, NE46-629, Cambridge, Massachusetts 02139, USA}} 
\affiliation{Department of Physics, Indian Institute of Science, Bangalore 560012, India}
\author{K. Hima Nagamanasa\footnote{Present address: Center for Soft and Living Matter, Institute for Basic Sciences, 50 UNIST-gil, Ulju-gun, Ulsan 44919, Republic of Korea (South Korea)}}
\affiliation{Chemistry and Physics of Materials Unit, Jawaharlal Nehru Centre for Advanced Scientific Research, Jakkur, Bangalore 560064, India}
\author{A. K. Sood}
\affiliation{Department of Physics, Indian Institute of Science, Bangalore 560012, India}
\affiliation{International Centre for Materials Science, Jawaharlal Nehru Centre for Advanced Scientific Research, Jakkur, Bangalore 560064, India}
\author{Rajesh Ganapathy$^*,$}
\affiliation{International Centre for Materials Science, Jawaharlal Nehru Centre for Advanced Scientific Research, Jakkur, Bangalore 560064, India}
\affiliation{Sheikh Saqr Laboratory, Jawaharlal Nehru Centre for Advanced Scientific Research
Jakkur, Bangalore, 560064, INDIA}

%\ead{$^1$ gokhales@mit.edu, $^3$ rajeshg@jncasr.ac.in}

\begin{abstract}
Elucidating the nature of the glass transition has been the holy grail of condensed matter physics and statistical mechanics for several decades. A phenomenological aspect that makes glass formation a conceptually formidable problem is that structural and dynamic correlations in glass-forming liquids are too subtle to be captured at the level of conventional two-point functions. As a consequence, a host of theoretical techniques, such as quenched amorphous configurations of particles, have been devised and employed in simulations and colloid experiments to gain insights into the mechanisms responsible for these elusive correlations. Very often, though, the analysis of spatio-temporal correlations is performed in the context of a single theoretical framework, and critical comparisons of microscopic predictions of competing theories are thereby lacking. Here, we address this issue by analysing the distribution of localized excitations, which are building blocks of relaxation as per the Dynamical Facilitation (DF) theory, in the presence of an amorphous wall, a construct motivated by the Random First-Order Transition theory (RFOT). We observe that spatial profiles of the concentration of excitations exhibit complex features such as non-monotonicity and oscillations. Moreover, the smoothly varying part of the concentration profile yields a length scale $\xi_c$, which we compare with a previously computed length scale $\xi_{dyn}$. Our results suggest a method to assess the role of dynamical facilitation in governing structural relaxation in glass-forming liquids. 
\end{abstract}

\date{\today}
\draft
\maketitle
\renewcommand{\thefootnote}

\section{Introduction}
Glass formation, the process whereby a flowing liquid transforms into a rigid amorphous solid remains one of the most actively pursued areas of research in condensed matter physics and statistical mechanics. Perhaps the most prominent yet confounding aspect of glass formation is that the drastic increase in viscosity of a liquid on approaching the glass transition is not accompanied by obvious changes in structure. The difficulty in identifying structural signatures of glass formation has led to the development of two parallel approaches aimed at explaining the observed phenomenology of the glass transition. The first approach advocates the notion that glass formation is associated with an underlying thermodynamic phase transition. The thermodynamic approach has been adopted by various theories such as the Adam Gibbs theory \cite{adam1965temperature}, the Random First-Order Transition theory (RFOT) \cite{kirkpatrick1989scaling,lubchenko2007theory,lubchenko2015theory} and theories based on geometric frustration \cite{tarjus2005frustration,tanaka1999two,royall2015role}. Although these theoretical formulations differ in significant way, they all postulate that structural relaxation proceeds via the reorganization of correlated domains that grow in size on approaching the glass transition. Thus, these approaches attribute the observed growth in viscosity or relaxation time to the existence of a growing structural length scale \cite{berthier2011theoretical,karmakar2014growing}. The other major approach to understanding glass formation is the Dynamical Facilitation (DF) theory \cite{garrahan2002geometrical,chandler2009dynamics}, which is purely kinetic in nature. Inspired by a class of spin models known as kinetically constrained models (KCMs) \cite{ritort2003glassy}, this approach claims that structural relaxation in glass-forming liquids is mediated by the concerted motion of spatio-temporally localized mobility carrying defects, called excitations. Since different formulations have distinct predictions for the temperature dependence of viscosity or relaxation time, one would expect experimental measurements of these quantities to provide unambiguous empirical evidence in favor of one theory or another. Unfortunately, viscosity data on molecular liquids even over fourteen orders of magnitude is insufficient to distinguish between the predictions of RFOT and DF, two of the most prominent theories of the glass transition \cite{biroli2013perspective}. It is therefore evident that in order to critically compare the predictions of competing theories, it is necessary to resort to a more microscopic spatio-temporally resolved analysis of relaxation dynamics of glass-forming liquids. 

A promising way to compare the predictions of competing theories is to examine the predictions of one theory in the  context of a framework developed for another. Here, as a concrete demonstration of this idea, we analyse the spatial distribution of the localized excitations of the DF theory in the presence of an amorphous wall, a construct designed to test ideas from RFOT \cite{kob2012non,nagamanasa2015direct}. By characterizing the spatial concentration profiles of excitations, we find evidence for non-monotonicity and oscillations in the concentration of excitations as a function of distance from the wall. More importantly, we define a new dynamic length scale $\xi_{c}$ associated with the distance over which the concentration of excitations saturates to its bulk value. We observe that $\xi_c$ exhibits non-monotonicity as a function of area fraction $\phi$, much like a previously computed length scale $\xi_{dyn}$ \cite{kob2012non,nagamanasa2015direct}. We postulate that a comparison of the dependence of the two length scales on $\phi$ may provide a means to assess the importance of dynamical facilitation in structural relaxation. 

\section{Experimental Methods}
We analyzed data from video microscopy experiments on a binary colloidal glass-former \cite{nagamanasa2015direct}. The glass-former consisted of polystyrene particles of diameters $\sigma_S =$ 1.05 $\mu$m and $\sigma_L = $ 1.4 $\mu$m. The particle size ratio $\sigma_L/\sigma_S =$ 1.3 and number ratio $N_L/N_S =$ 1.23 adequately suppressed crystallization over the experimental duration. The samples were loaded into a wedge-shaped cell and the area fraction $\phi$ was tuned via controlled sedimentation of particles to the monolayer-thick region of the wedge. We observed a waiting time of about 8-10 hours before freezing the wall to ensure sample equilibration \cite{nagamanasa2015direct}. This waiting time is several times larger than the structural relaxation time $\tau_{\alpha}$ for $\phi \leq$ 0.76, confirming that these samples were equilibrated. For $\phi =$ 0.79, the system did not relax fully over the experimental duration of 1.5 hours. However, the waiting time is about six times larger than the duration of the experiment. Samples were imaged using a Leica DMI 6000B optical microscope with a $\times$100 objective (Plan apochromat, NA 1.4, 98 oil immersion) and images were captured at frame rates ranging from 3.3 fps to 5 fps for 1 to 1.5 hours, depending on the area fraction. The size of the field of view was 44$\sigma$ $\times$ 33$\sigma$, where $\sigma = (\sigma_S+\sigma_L)/2$. An amorphous wall was created by simultaneously trapping $\approx$ 100 colloids using holographic optical tweezers. The holographic optical tweezers set-up comprised of a linearly polarized constant power (800mW) CW laser (Spectra-Physics, 1064 nm) and optical traps were created using a Spatial Light Modulator (512 $\times$ 512, 100 fps refresh rate, Boulder Nonlinear Systems). Standard Matlab algorithms \cite{crocker1996methods} were used to construct particle trajectories and subsequent analysis was performed using codes developed in-house.

\section{Results and Discussion}
\begin{figure}[tbp]
\centering
\includegraphics[width=0.7\columnwidth]{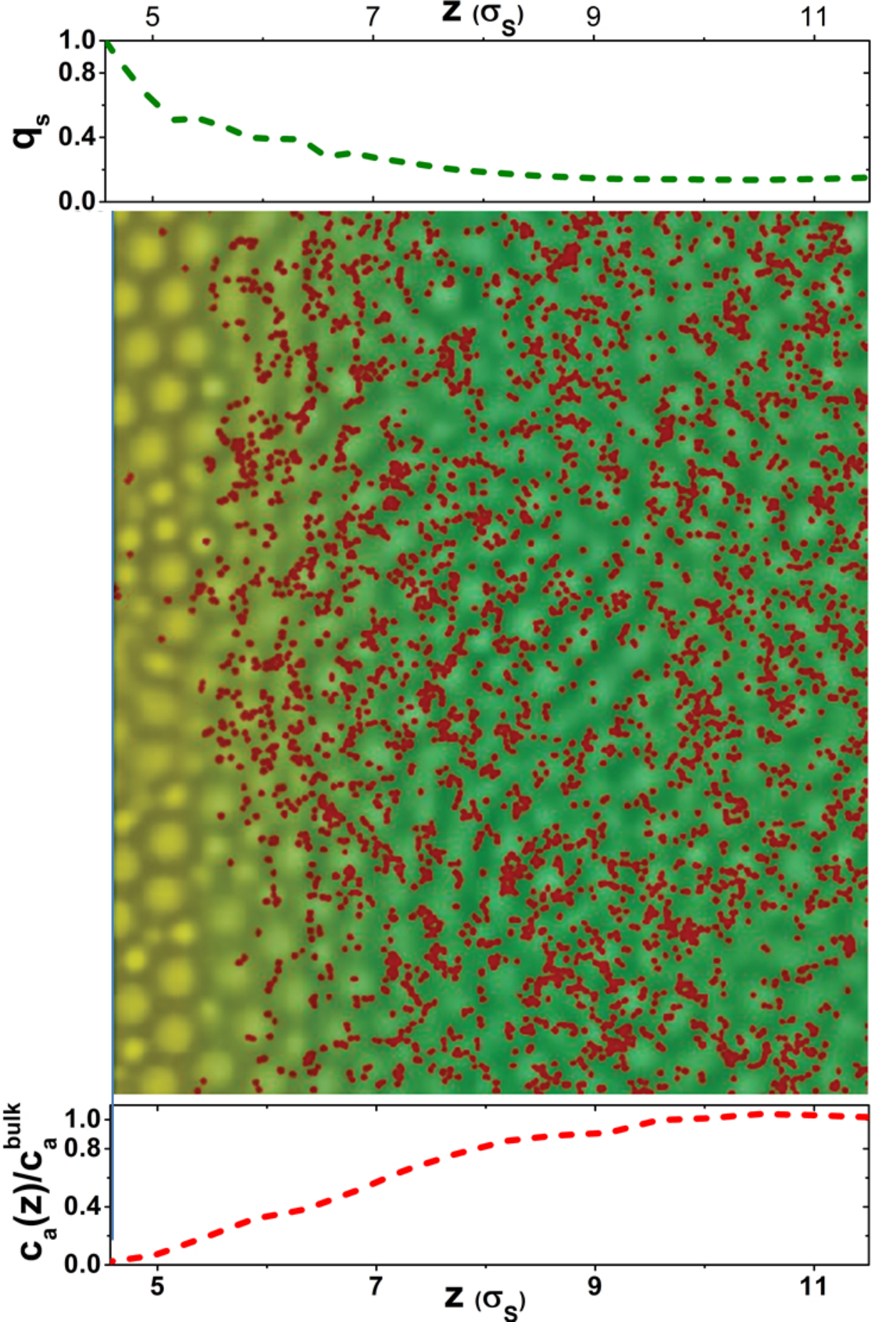}
\caption{Variation of the relaxation time and $c_a$ with $z$. The background image has been generated by averaging individual frames for $ \phi =$ 0.74 over a time interval of 180 s, which corresponds to three times the structural relaxation time $\tau_{\alpha}$. The yellow-green colormap superimposed on the background image corresponds to the time-averaged self-overlap $q_s = \langle q_s(t) \rangle_{t'}$ (See \cite{nagamanasa2015direct}), where $t'$ denotes a time interval of 40 s centred on $\tau_{\alpha}$. The small red spheres correspond to excitations of size $a =$ 0.23$\sigma_S$. The dotted green and red curves show the variation of $q_s$ and $c_a(z)$ with $z$, respectively.}
\label{Figure1}
\end{figure}

The amorphous wall is one of a class of pinning geometries that have been primarily used to extract a static point-to-set length scale \cite{berthier2012static}. It has recently risen to prominence since it was instrumental in extracting a dynamic length scale $\xi_{dyn}$ that exhibits a non-monotonic dependence on temperature \cite{kob2012non} or area fraction $\phi$ \cite{nagamanasa2015direct}. In physical terms, $\xi_{dyn}$ is a measure of the distance over which the structural relaxation time $\tau_{\alpha}$ is influenced by the presence of the amorphous wall. Broadly speaking, the presence of the wall induces a spatial variation in the relaxation time and $\xi_{dyn}$ characterizes the nature of this variation. The non-monotonicity observed in simulations as well as experiments likely arises from a change in the morphology of cooperatively rearranging regions from string-like to compact form and is consistent with the prediction of RFOT \cite{stevenson2006shapes}. In the present study, our primary aim is to examine whether these experimental and numerical findings can be reconciled within the DF theory as well. The DF theory claims that immobile regions in a glass-forming liquid can become mobile only if they are in the vicinity of a mobility-carrying excitation. This implies that the relaxation time has a one-to-one correspondence with the concentration of excitations: the larger the concentration of excitations, the faster the system relaxes. If this assumption of the DF theory holds in the presence of an amorphous wall, the dependence of $\tau_{\alpha}$ with $z$ should be mirrored by the concentration of excitations (Fig. \ref{Figure1}). In other words, the concentration of excitations should be consistent with $\tau_{\alpha}(z)$ for all $z$, and hence, the length scale $\xi_c$ characterizing the concentration profile should exhibit the same dependence on $\phi$ as $\xi_{dyn}$. 

\begin{figure}[tbp]
\centering
\includegraphics[width=0.8\columnwidth]{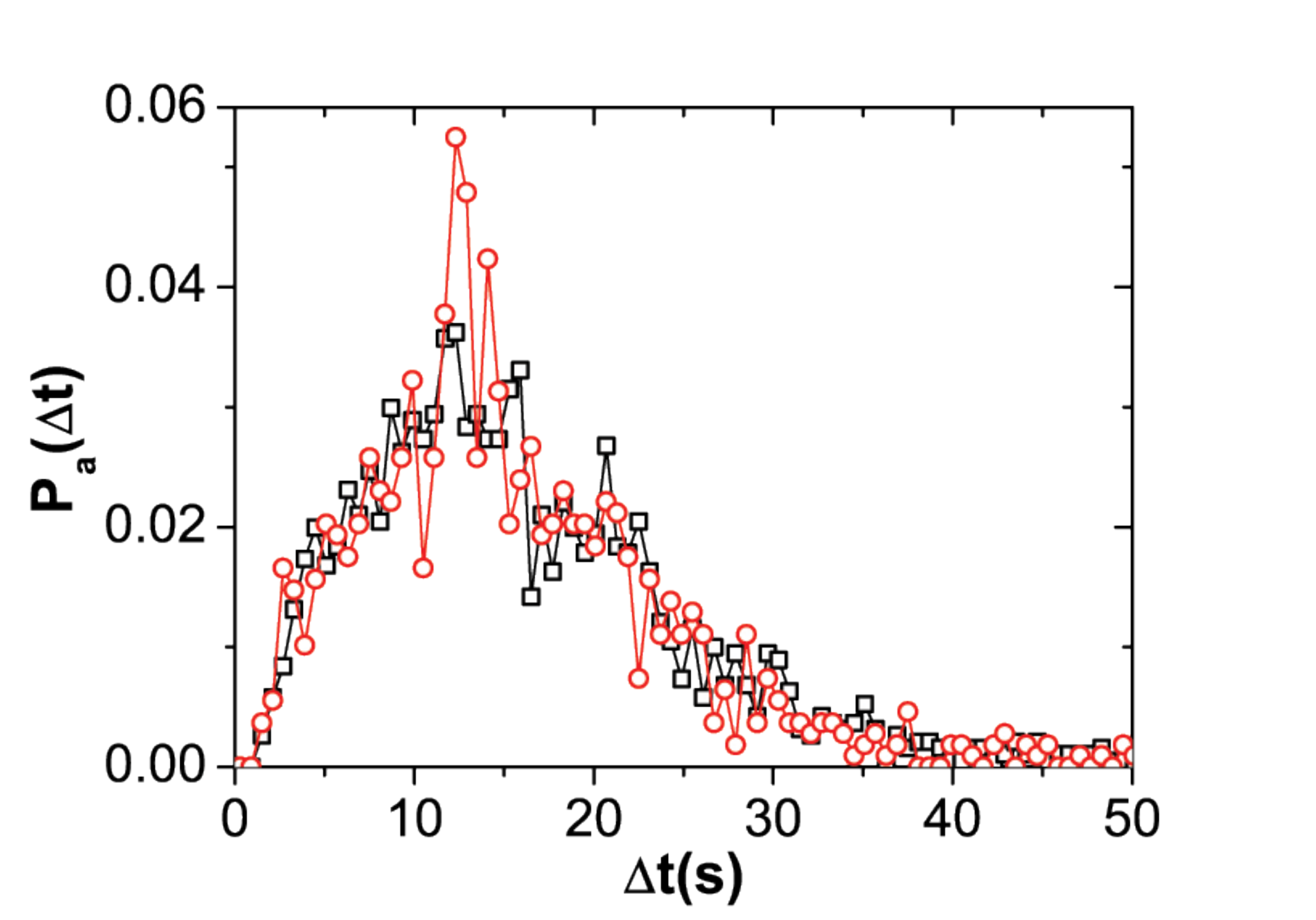}
\caption{Instanton time distribution $P_a(\Delta t)$ for $a =$ 0.46$\sigma_S$ and $\phi =$ 0.74 for two halves of the field of view. $P_a(\Delta t)$ for the half that contains the amorphous wall is shown by ({\color{red} $ \circ$}) and that for the remaining half is shown by ({\color{black} $ \square$}).}
\label{Figure2}
\end{figure}

With this objective in mind, we analysed the excitation concentration profiles for various $\phi$. To identify excitations, we followed the procedure developed in \cite{keys2011excitations}. Since the method has been described in detail in our previous works \cite{gokhale2014growing,gokhale2016localized}, we outline it very briefly here. A particle is associated with an excitation of size $a$ and instanton time duration $\Delta t$, if it undergoes a displacement of magnitude $a$ over time $\Delta t$ and persists in its initial as well as final state for at least $\Delta t$ \cite{keys2011excitations}. Excitations were identified by first coarse-graining particle trajectories over a suitable time window \cite{mishra2014dynamical}, and then computing the functional
\begin{equation}
h_{i}(t,t_{a};a) = \prod\limits_{t' = t_{a}/2 - \Delta t}^{t_{a}/2} \theta(|\bar{\textbf{r}}_{i}(t+t')-\bar{\textbf{r}}_{i}(t-t')| - a)
\end{equation}
where, $\theta(x)$ is the Heaviside step function \cite{keys2011excitations,gokhale2014growing}. $h_{i}(t,t_{a};a) =$ 1 whenever the trajectory is associated with an excitation and 0 whenever it is not. Here, $t_a$, known as the commitment time is typically about three times the mean instanton time \cite{keys2011excitations}. We performed this analysis of excitations of two different excitation sizes $a =$ 0.23$\sigma_S$ and $a =$ 0.46$\sigma_S$. Excitation dynamics is hierarchical, the results are expected to be similar for other sizes as well \cite{keys2011excitations}. To ensure that the instanton time distribution $P(\Delta t)$ is unaffected by the presence of the wall, we computed this distribution for two halves of the field of view, one of which contained the wall. We observe that the two distributions overlap completely, demonstrating that the nature of excitations is not influenced by the presence of the wall (Fig. \ref{Figure2}). 

\begin{figure}[tbp]
\centering
\includegraphics[width=1\columnwidth]{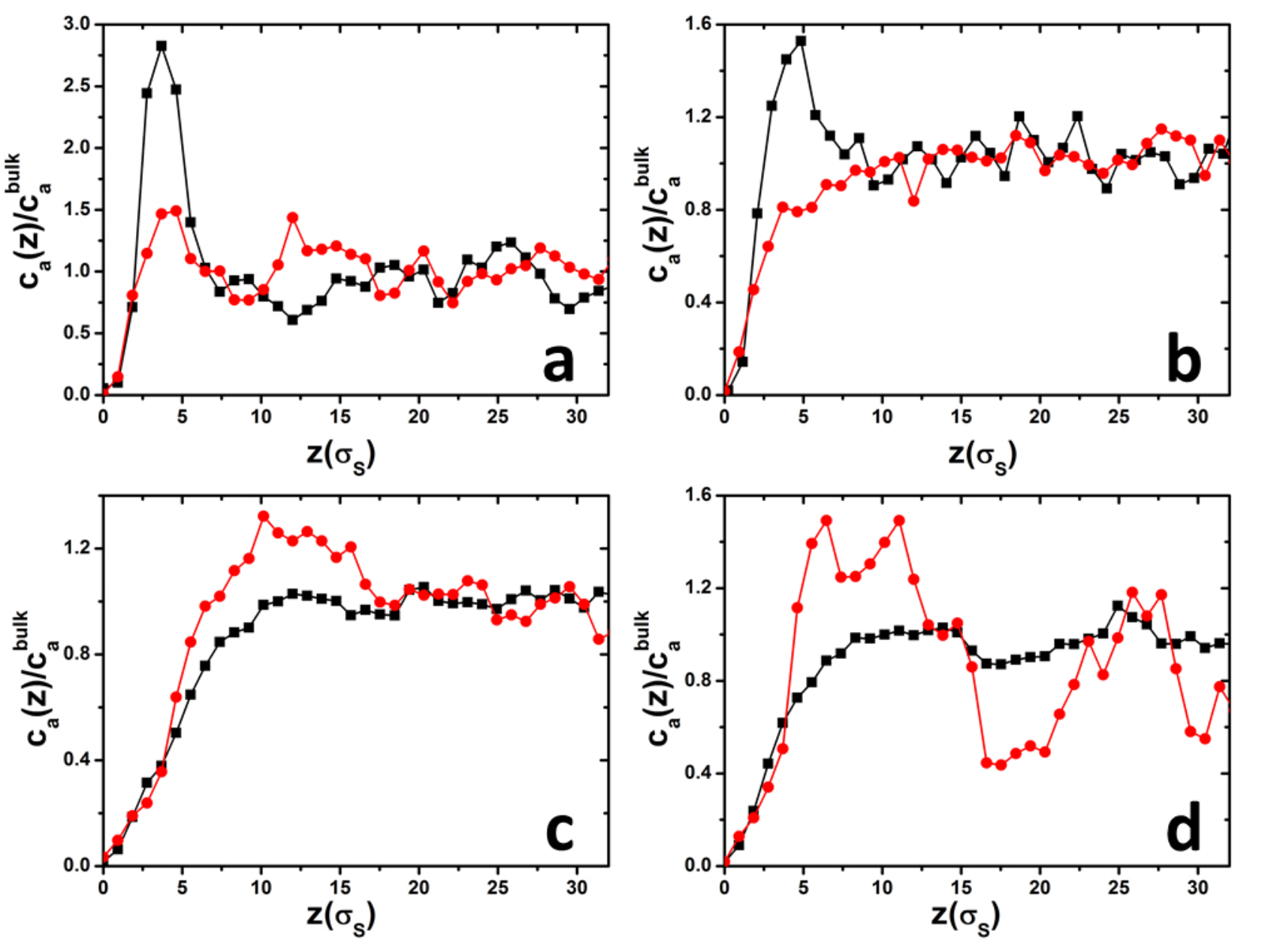}
\caption{The excitation concentration profiles $c_a(z)$ normalized by their respective bulk values $c_a^{bulk}$ for $a =$ 0.23$\sigma_S$ ({\color{black} $ \blacksquare$}) and $a =$ 0.46$\sigma_S$ ({\color{red} $ \bullet$}) for $\phi =$ 0.68 (a), $\phi =$ 0.71 (b), $\phi =$ 0.74 (c) and $\phi =$ 0.75 (d).}
\label{Figure3}
\end{figure}

Intuitively, $c_a$ should be zero at the wall, i.e. $c_a(z=0)=0$, since relaxation does not occur at the wall. Further, at large distances from the wall, $c_a$ is expected to reach its bulk value $c_{a}^{bulk} = c_a(z\rightarrow\infty)$. Naively, therefore, one expects $c_a(z)$ to be a monotonically increasing function that interpolates smoothly between these two limits. In reality, the profiles $c_a(z)$ are much more complex, exhibiting non-monotonicity and even oscillations as a function of distance from the wall (Fig. \ref{Figure3}). For $\phi \leq$ 0.71, we observe a strong peak for $a =$ 0.23$\sigma_S$ (Fig. \ref{Figure3}a). We believe that this enhancement occurs because at these low values of $a$ and $\phi$ the time scales of particle caging and cage-breaking are not well separated. The determination of $c_a$ is accurate if particle trajectories can be neatly divided into quiescent regimes corresponding to vibrational motion within cages formed by nearest neighbors and rare, sporadic events in which particles escape from one cage and get trapped in another \cite{keys2011excitations}. Since $\phi =$ 0.68 is very close to the onset of caging, particle trajectories are nearly continuous and hence, fewer excitations are identified. Close to the wall, however, the relaxation time increases and caging becomes more prominent, which leads to a larger value of $c_{a}$ as compared to the bulk. Finally, $c_a$ drops down to zero at the wall since no relaxation is possible there, which leads to a maximum in the concentration profile. Since the separation of timescales becomes more pronounced for larger $a$ and $\phi$, we expect the peak to diminish in amplitude with increasing $a$ as well as $\phi$. Fig. \ref{Figure3}a-c shows that this is indeed the case. Rather counter-intuitively, however, for intermediate values of $\phi$ (Fig. \ref{Figure3}c-d) the overshoot in $c_a(z)$ is larger for $a =$ 0.46$\sigma_S$ than for $a =$ 0.23$\sigma_S$. Since the concentration of excitations decreases with increasing $\phi$, this observation may also be an effect of increasing noise. In fact, at the two largest $\phi$ studied, we were unable to explore the $a$ dependence of the concentration profiles due to poor statistics. 

\begin{figure}[tbp]
\centering
\includegraphics[width=1\columnwidth]{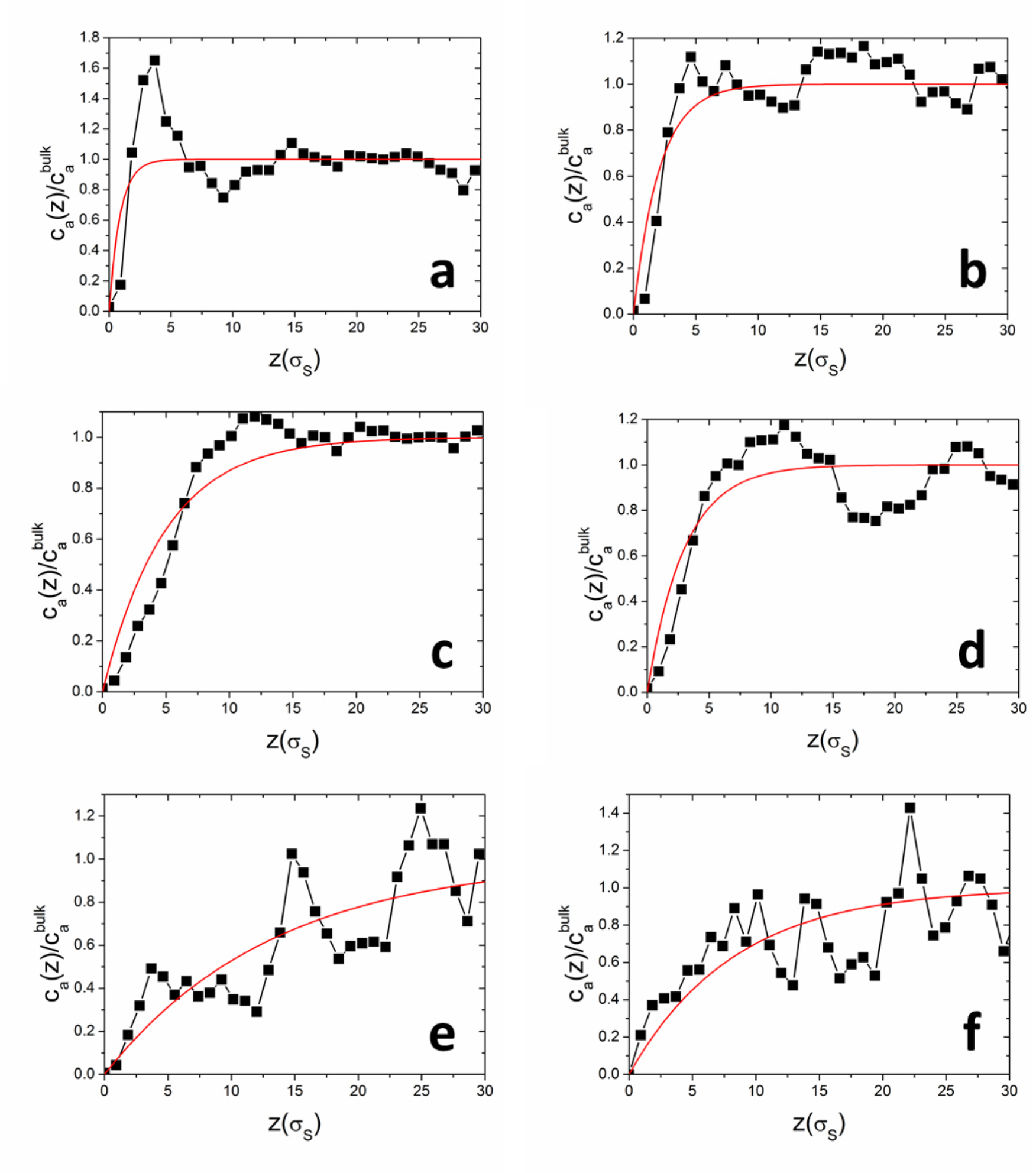}
\caption{Spatial concentration profiles of excitations $c_a(z)$ averaged over excitation sizes ranging from $a =$ 0.23$\sigma_S$ to $a =$ 0.61$\sigma_S$ and normalized by their respective bulk values $c_a^{bulk}$ for $\phi =$ 0.68 (a), $\phi =$ 0.71 (b), $\phi =$ 0.74 (c), $\phi =$ 0.75 (d), $\phi =$ 0.76 (e) and $\phi =$ 0.79 (f). The $c_a^{bulk}$ values have been computed by averaging $c_a(z)$ over a window of 5$\sigma_S$-10$\sigma_S$ towards the end of the profile, i.e. for $z \geq$ 20$\sigma_S$. The red curves are empirical fits of the form $(1-\textrm{exp}(-x/\xi_c))$ from which we extract the length scale $\xi_c$.}
\label{Figure4}
\end{figure}

\begin{figure}[tbp]
\centering
\includegraphics[width=1\columnwidth]{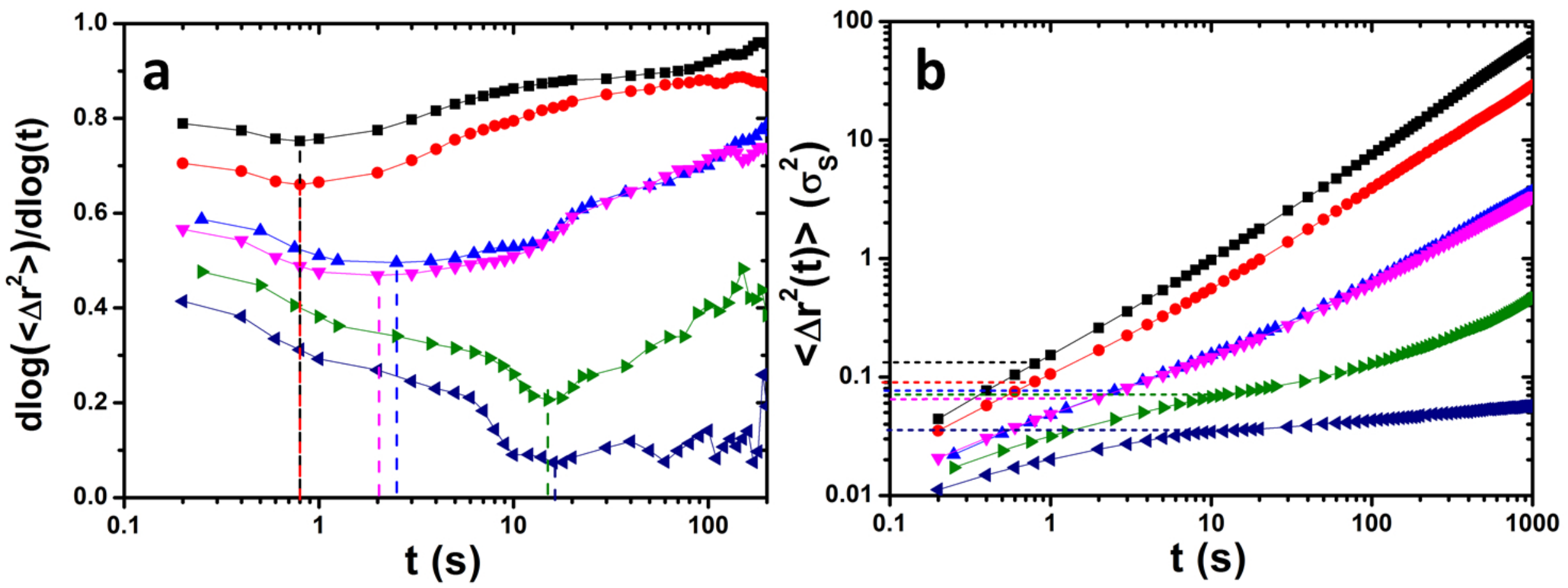}
\caption{a) Determination of the beta relaxation time $\tau_{\beta}$. $d\log(\langle \Delta r^2(t) \rangle)/d\log(t)$ for $\phi =$ 0.68 ({\color{black} $ \blacksquare$}), $\phi =$ 0.71 ({\color{red} $ \bullet$}), $\phi =$ 0.74 ({\color{blue} $ \blacktriangle$}), $\phi =$ 0.75 ({\color{magenta} $ \blacktriangledown$}), $\phi =$ 0.76 ({\color{green!50!black} $ \blacktriangleright$}) and $\phi =$ 0.79 ({\color{blue!50!black} $ \blacktriangleleft$}). The dashed vertical lines denote corresponding values of $\tau_{\beta}$ b) The mean squared displacement $\langle \Delta r^2(t) \rangle$ for various $\phi$. The colors and symbols are identical to those in (a). The dashed horizontal lines denote the square of the cage size $R_c$, defined as $R_c = \sqrt{\langle \Delta r^2(\tau_{\beta}) \rangle}$, evaluated using the values of $\tau_{\beta}$ obtained from (a).}
\label{Figure5}
\end{figure}

\begin{figure}[tbp]
\centering
\includegraphics[width=1\columnwidth]{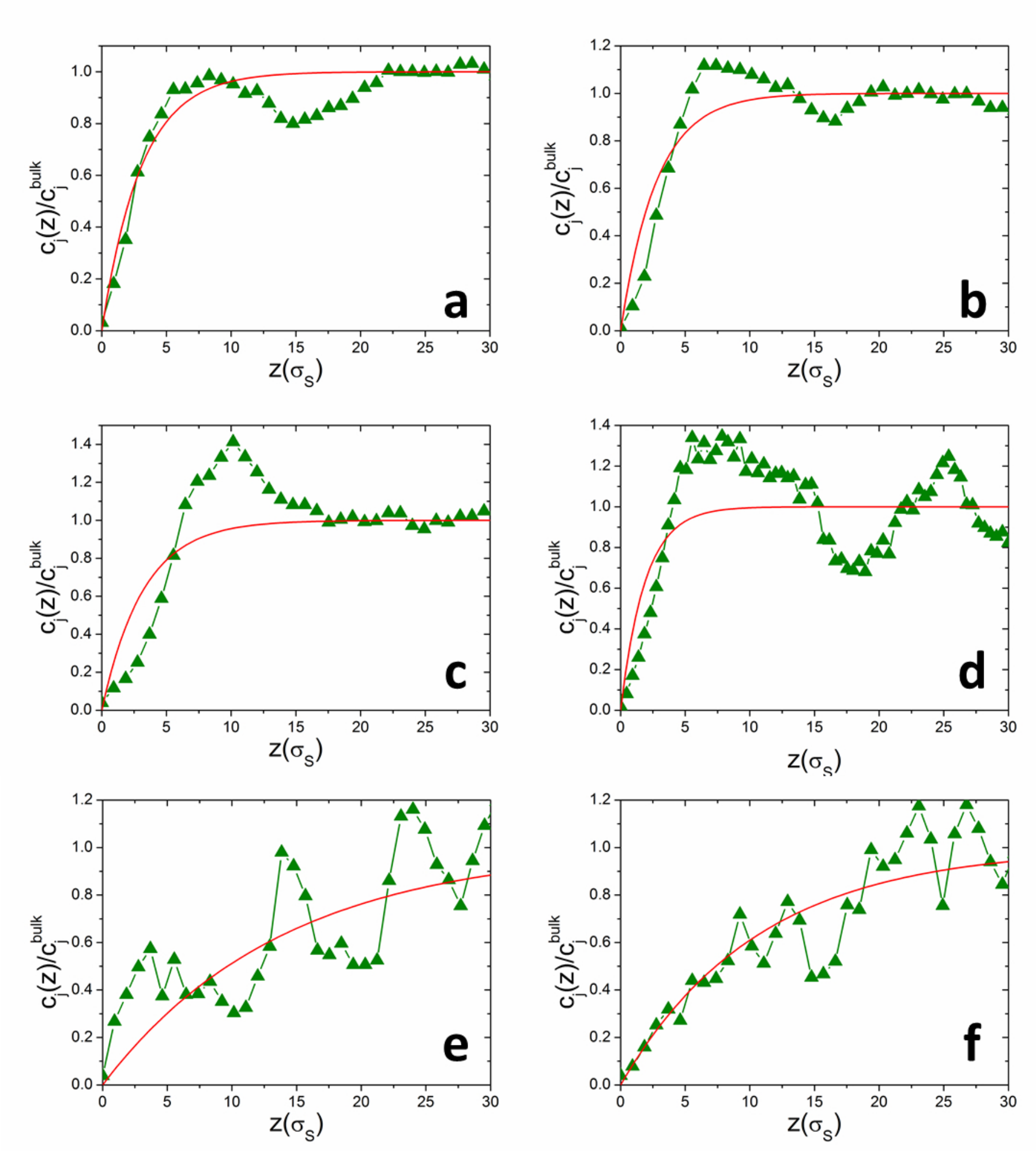}
\caption{Spatial concentration profiles of cage jumps $c_j(z)$ normalized by their respective bulk values $c_j^{bulk}$ for $\phi =$ 0.68 (a), $\phi =$ 0.71 (b), $\phi =$ 0.74 (c), $\phi =$ 0.75 (d), $\phi =$ 0.76 (e) and $\phi =$ 0.79 (f). The $c_j^{bulk}$ values have been computed by averaging $c_j(z)$ over a window of 5$\sigma_S$-10$\sigma_S$ towards the end of the profile, i.e. for $z \geq$ 20$\sigma_S$. The red curves are empirical fits of the form $(1-\textrm{exp}(-z/\xi_j))$ from which we extract the cage jump length scale $\xi_j$.}
\label{Figure6}
\end{figure}

For $\phi =$ 0.75, the concentration profile suggests the presence of oscillations (Fig. \ref{Figure3}d). For $\phi =$ 0.76, we find that pronounced oscillatory features can be observed even when the concentration profile is averaged over excitation sizes ranging from $a =$ 0.23$\sigma_S$ to $a =$ 0.61$\sigma_S$ (Fig. \ref{Figure4}e). A close inspection of the concentration profiles in Fig. \ref{Figure3} as well as Fig. \ref{Figure4} shows that oscillations are present, albeit less pronounced, even at other area fractions. To ensure that these oscillations are not an artefact of the protocol used for identifying excitations, we defined excitations using the cage jump analysis developed by Biroli and co-workers \cite{candelier2009building,candelier2010spatiotemporal}. The cage jump analysis identifies excitations as cage-breaking events that divide particle trajectories into intervals of rattling within distinct cages. For clarity, we shall henceforth refer to excitations identified using the cage jump analysis as `cage jumps'. To identify cage jumps, we considered a particle trajectory $S(t)$ of total duration $T$ and divided it into two sub-trajectories $S_1(t_1)$ and $S_2(t_2)$ at an arbitrarily chosen time instant $t_c$, such that $t_1 \in [0,t_c]$ and $t_2 \in [t_c,T]$. Next, we quantified the spatial separation between the two sub-trajectories, $p(t_c)$, as follows
\begin{equation}
p(t_c) = \xi(t_c)\sqrt{\langle d_1(t_2)^2 \rangle_{t_2 \in S_2} \langle d_2(t_1)^2 \rangle_{t_1 \in S_1}}
\end{equation}
Here, $\xi(t_c) = \sqrt{(t_c/T)(1 - t_c/T)}$ and $d_i(t_j)$ is the distance between the particle's position at time $t_j$ from the centre of mass of the sub-trajectory $S_i$. A cage jump is said to occur at the time $t_c$ at which $p(t_c)$ is maximal, or in other words, the two subtrajectories $S_1$ and $S_2$ are maximally separated in space. The procedure is repeated recursively until $p_{max}(t_c) < R_c$, where $R_c$ is the cage size. The cage size of colloidal glass-forming liquids is known to decrease with $\phi$ \cite{weeks2002properties}. In order to determine the cage size, we first computed the caging, or beta relaxation time $\tau_{\beta}$ from the mean squared displacement of particles $\langle \Delta r^2(t) \rangle$. Specifically, we defined $\tau_{\beta}$ as the time that minimizes $d\log(\langle \Delta r^2(t) \rangle)/d\log(t)$ (Fig. \ref{Figure5}a). The cage size was then extracted from $\langle \Delta r^2(t) \rangle$, using the definition $R_c = \sqrt{\langle \Delta r^2(\tau_{\beta}) \rangle}$ (Fig. \ref{Figure5}b). While the procedures for identifying excitations and cage jumps are clearly different, they are similar in spirit in that both of them rely on the separation between the time scales associated with cage rattling and cage escapes \cite{keys2011excitations}. We observe that the cage jump concentration profiles, $c_j(z)$ (Fig. \ref{Figure6}), bear a strong resemblance to the excitation concentration profiles averaged over $a$ (Fig. \ref{Figure4}), particularly for $\phi \geq$ 0.74, where the separation of time scales associated with cage rattling and cage escapes is well-defined. In particular, we observe that the pronounced oscillations for $\phi =$ 0.76 (Fig. \ref{Figure4}e) are also present in the corresponding cage jump concentration profile (Fig. \ref{Figure6}e). This shows conclusively that spatial oscillations are an intrinsic dynamical consequence of an amorphous wall and not a mere artefact. In previous work, we have shown that $\phi =$ 0.76 coincides with the point at which the system exhibits non-monotonicity in dynamic correlations, which is accompanied by the change in morphology of CRRs \cite{nagamanasa2015direct}. Whether or not these results have any bearing on the observed oscillations in $c_a(z)$ is worth investigating in future studies. It might also be worthwhile to examine whether the oscillations arise due to the interaction of mobility surges emanating from excitations \cite{keys2011excitations} with the static amorphous wall. 

While the complex features of $c_a(z)$ and $c_j(z)$, such as non-monotonicity and oscillations are interesting, the underlying monotonically increasing average profile that characterizes $c_a(z)$ also yields valuable information about structural relaxation. Crucially, it suggests a potential way of gauging the importance of facilitation as a relaxation process. The most interesting feature of the monotonic part of the profile is the characteristic length scale $\xi_c$ over which $c_a(z)$ saturates to its bulk value $c_{a}^{bulk}$. We extract this length scale from empirical fits of the form $c_a(z)/c_{a}^{bulk} = 1 - \textrm{exp}(-z/\xi_c)$ (Fig. \ref{Figure4}). Since excitations of different sizes differ only in their formation energy \cite{keys2011excitations}, one expects $\xi_c$ to be independent of $a$. and the concentration profiles shown in Fig. \ref{Figure3} are consistent with this expectation in that the distance over which $c_a(z)$ for $a =$ 0.23$\sigma_S$ and $a =$ 0.46$\sigma_S$ saturate to their bulk values is approximately equal. However, the complex features of $c_a(z)$ described in preceding paragraphs may influence the estimate of $\xi_c$. To minimize this influence, we have extracted the length scale $\xi_c$ from the excitation concentration profiles averaged over $a$ (Fig. \ref{Figure4}). We have also extracted a length scale $\xi_j$ from the cage jump concentration profiles $c_j(z)$, by fitting the form $c_j(z)/c_{j}^{bulk} = 1 - \textrm{exp}(-z/\xi_j)$. 

\begin{figure}[tbp]
\centering
\includegraphics[width=0.7\columnwidth]{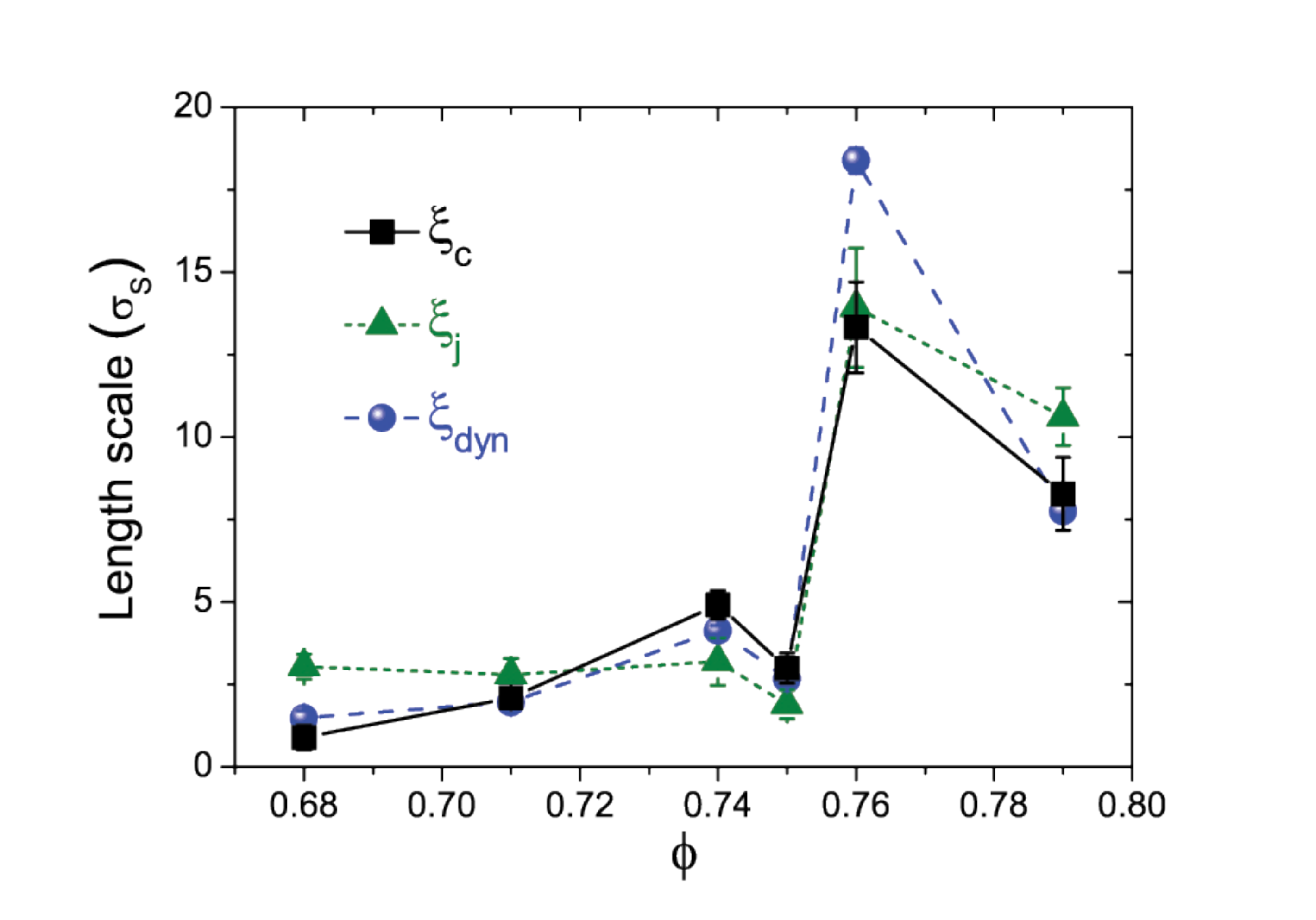}
\caption{The variation of dynamic length scales $\xi_{dyn}$ ({\color{blue} $ \bullet$}), taken from \cite{nagamanasa2015direct}, $\xi_c$ ({\color{black} $ \blacksquare$}) and $\xi_j$ ({\color{green!50!black} $ \blacktriangle$}) with $\phi$. The errors bars correspond to the error on the fitting parameter. Since we have not averaged the data over multiple realizations of the amorphous wall, the actual errors in estimating the length scales are larger than those indicated by the error bars.}
\label{Figure7}
\end{figure}

Figure \ref{Figure7} compares the evolution of these length scales on approaching the glass transition, with that of the dynamic length scale $\xi_{dyn}$ obtained from the variation of the relaxation time with distance from the amorphous wall \cite{nagamanasa2015direct}. First, we observe that $\xi_c$ and $\xi_j$ appear to be in reasonable agreement with each other, once again suggesting that they are associated with similar dynamical events. In the context of ascertaining the relevance of facilitation as a relaxation mechanism, it is more instructive to compare $\xi_c$ with $\xi_{dyn}$. As mentioned earlier, according to the DF theory, the relaxation time has a one-to-one correspondence with the concentration of excitations and hence, $\xi_c$ and $\xi_{dyn}$ must exhibit identical scaling on approaching the glass transition. We observe that in general, the two length scales are comparable at all area fractions except $\phi =$ 0.76 and in particular $\xi_c$ also exhibits non-monotonicity as a function of $\phi$. Recent experimental results based on the analysis of spatial organization of excitations within clusters of mobile particles suggests that facilitation diminishes in importance for $\phi >$ 0.75 \cite{gokhale2016localized}. Thus, one interpretation of the data in Fig. \ref{Figure7} is that the decoupling of $\xi_c$ and $\xi_{dyn}$ at $\phi =$ 0.76 is consistent with the diminishing role of facilitation. Such a conclusion is also supported by the $\phi$ dependence of the mobility transfer function \cite{nagamanasa2015direct}. However, we note that in our experiments, it was not feasible to perform disorder averaging, i.e. averaging over multiple realizations of the amorphous wall. Further, the data at high $\phi$ have larger errors, since excitations become rarer on approaching the glass transition. Finally, for the largest area fraction, $\phi =$ 0.79, it was not possible to determine conclusively whether the sample was in equilibrium. Since the distribution of distances between excitations depends on sample equilibration \cite{keys2015using}, the equilibration time may influence the excitation concentration profiles. Given this uncertainty in determining the concentration profiles $c_a(z)$, one must exercise caution while interpreting trends in $\xi_c$ at large $\phi$. We therefore hope that our findings are tested extensively and rigorously in future studies in order to determine whether or not dynamical facilitation dominates structural relaxation over the dynamical regime accessible to experiments as well as simulations. 

\section{Conclusions}
By analyzing data from colloid experiments, we have characterized the variation of the concentration of localized excitations with distance from an amorphous wall. Contrary to the naive expectation of a monotonic increase followed by saturation, we find that the concentration profiles $c_a(z)$ exhibit complex non-monotonic and even oscillatory features (Fig. \ref{Figure4}). We confirmed the existence of spatial oscillations in the concentration profiles by identifying analogues of excitations known as cage jumps and analysing their distribution $c_j(z)$ as a function of distance from the wall (Fig. \ref{Figure6}). Crucially, we have extracted new length scales $\xi_c$ and $\xi_j$ from $c_a(z)$ and $c_j(z)$, respectively, and compared them to a previously quantified dynamic length scale $\xi_{dyn}$. Further experiments and simulations are necessary to determine whether the discrepancy between the two length scales at $\phi =$ 0.76 is consistent with the recently observed crossover from facilitation to collective hopping observed in colloidal glass-formers \cite{gokhale2016localized}. In a broader context, the length scale $\xi_c$ may serve as a generic diagnostic tool to infer the relevance of facilitation in a variety of real as well as simulated glass-forming liquids. 

\section*{Acknowledgments}
R.G. thanks the International Centre for Materials Science (ICMS) and the Sheikh Saqr Laboratory (SSL), Jawaharlal Nehru Centre for Advanced Scientific Research (JNCASR) for financial support and A.K.S. thanks Department of Science and Technology (DST), India for support under J.C. Bose Fellowship. S.G. thanks DST, India, for financial support. 

\section*{Corresponding Authors}
\noindent Shreyas Gokhale\\
Email: gokhales@mit.edu\\
\noindent Rajesh Ganapathy\\
Email: rajeshg@jncasr.ac.in\\

\newpage

\bibliography{references}
\bibliographystyle{apsrev4-1}

\newpage

\end{document}